\documentclass[onecolumn,12pt]{IEEEtran}
\usepackage{amsmath}
\usepackage{amsthm}
\usepackage{amsfonts}
\usepackage{amssymb}
\usepackage{mathrsfs}
\usepackage{cases}
\usepackage{cite}
\usepackage{graphicx}
\usepackage{mathrsfs}
\usepackage{subfigure}
\usepackage{epstopdf}
\usepackage{color}

\usepackage{enumitem}
\usepackage{setspace}
\usepackage{bm}
\usepackage{caption}
\usepackage{algorithm}
\usepackage{algorithmic}
\usepackage{multirow}
\usepackage{tabularx}

\theoremstyle{remark}

\begin{document}
\baselineskip 4.2ex
\title{Joint Space-Time Sparsity Based Jamming Detection for Mission-Critical mMTC Networks}
\author{Shao-Di Wang, Hui-Ming Wang,~\IEEEmembership{Senior Member,~IEEE},\\ Zhetao Li,~\IEEEmembership{Senior Member,~IEE}, and Victor C. M. Leung,~\IEEEmembership{Fellow,~IEEE}
\thanks{\scriptsize
This article was presented in part at the IEEE Global Communications Conference, Rio de Janeiro, Brazil, December 2022 [1]. (\emph{Corresponding author: Hui-Ming Wang}.)}
\thanks{\scriptsize
S.-D. Wang and H.-M. Wang are with the School of Information and Communication Engineering, and also with the Ministry of Education Key Lab for Intelligent Networks and Network Security, Xi’an Jiaotong University, Xi’an, 710049, Shaanxi, China (e-mail: xjtuwsd@stu.xjtu.edu.cn; xjbswhm@gmail.com).
}
\thanks{\scriptsize
Z. Li is with the Key Laboratory of Hunnan Province for Internet of Things and Information Security, Hunan International Scientific and Technological Cooperation Base of Intelligent Network, School of Computer Science, Xiangtan University, Xiangtan 411105, China (e-mail: liztchina@hotmail.com).
}
\thanks{\scriptsize
V. C. M. Leung is with the College of Computer Science and Software Engineering, Shenzhen University, Shenzhen 518060, China, and also with the Department of Electrical and Computer Engineering, The University of British Columbia, Vancouver, BC V6T 1Z4, Canada (e-mail: vleung@ieee.org).
}
}

\IEEEtitleabstractindextext{
\begin{abstract}
For mission-critical massive machine-type communications (mMTC) applications, the messages are required to be delivered in real-time. However, due to the weak security protection capabilities of the low-cost and low-complexity machine-type devices, active jamming attack in the uplink access is a serious threat. Uplink access jamming (UAJ) can increase the number of dropped/retransmitted packets and restrict or prevent the normal device access. To tackle this vital and challenging problem, we propose a novel UAJ detection method based on the joint space-time sparsity (JSTS). Our key insight is that the JSTS-based feature will be significantly impacted if UAJ happens, since only a small fraction of the devices are active and the traffic pattern for each device is sporadic in the normal state. Unlike the existing detection methods under batch mode (i.e., all sample observations are collected before making a decision), the JSTS-based detection is performed in a sequential manner by processing the received signals one by one, which can detect UAJ as quickly as possible. Moreover, the proposed JSTS-based method does not rely on the prior knowledge of the attackers, since it only cares the abrupt change in the JSTS-based feature on each frame. Numerical results evaluate and confirm the effectiveness of our method.

\end{abstract}
}
\maketitle

\IEEEdisplaynontitleabstractindextext
\IEEEpeerreviewmaketitle

\section{Introduction}

Mission-critical massive machine-type communications (mMTC) applications are emerging as an important service in the fifth generation (5G) for delay-sensitive traffics and have been drawing increasing attention recently [2], [3]. This kind of applications have stringent access latency and reliability requirements to facilitate mission-critical services. For instance, in smart manufacturing lines, the control system needs to monitor the condition of the manufacturing lines and makes real-time decisions, e.g., the latency is within several milliseconds [4]. To accommodate the challenging requirements of such applications, the grant-free random access scheme is proposed for 5G new radio in the third Generation Partnership Project Release 17 [5]. In grant-free scheme, each active device directly transmits its unique preamble sequence to the base station (BS), which simplifies the access procedure by directly delivering data without scheduling [6].

However, for mission-critical mMTC applications with grant-free scheme, active jamming attacks are a big challenge [7], especially in the uplink access. In this paper, we consider the uplink access jamming (referred to as UAJ) in the mission-critical mMTC, where a small number of attackers aim to affect the performance of such applications with low latency requirement by jamming. UAJ not only severely affects the availability of this kind of applications but also is hard to detect. On one hand, UAJ can lead to additional access collisions and activity patterns under this kind of applications with the feature of sporadic traffic in massive access, and increase the number of dropped/retransmitted packets rapidly by only a small number of attackers, which incur longer packet transmission delay. On the other hand, due to the weak security protection capabilities of low-cost low-power machine-type devices (MTDs), UAJ attackers can easily obtain the preamble sequences by eavesdropping and masquerade as the legitimate MTDs, and even some legitimate MTDs can be hijacked by UAJ attackers. Under the cover of the legitimate identities, a small number of activated UAJ attackers are difficult to detect. 

Therefore, a timely and reliable detection of UAJ is a critical issue to be addressed for mission-critical mMTC applications in order to ensure countermeasures can be timely taken, such as adaptive array beamforming [8], interference cancellation techniques [9], and interference alignment techniques [10].

\subsection{Related Literatures}

Most existing jamming detection can be performed by a \emph{statistic feature (SF) recoginition and classification approach}, and different detection methods use different statistics for decision making. These statistic features can be classified into the following two categories: \emph{1) statistical features of upper layer [11]-[14]}; \emph{2) statistical features of physical layer [15]-[19]}.

\emph{1) Statistical features of upper layer [11]-[14]:}  In [11], the effective channel utilization is calculated and used as a statistic to detect jamming attacks. This statistic is essentially the sum of channel access activities and signal strengths. In a dynamic radio channel allocation model, this strategy has been proved to be effective. In [12]-[13], the authors proposed to exploit the packet delivery ratio to detect the attacks, and activity patterns were further utilized to identify malicious attacker. In [14], the authors proposed to extract the statistical feature of the network throughput from the upper layer for detecting the attacks. As a result, a jamming attack is detected when the percentage exceeds a certain threshold.

\emph{2) Statistical features of physical layer [15]-[19]:}  In [15], the authors proposed an energy detector based method. This method is based on the fact that when jamming attacks occur, the received energy differs significantly from a predesigned threshold. By extracting the subspace dimension of the signal covariance matrix, a method based on subspace dimension was proposed in [16] to detect a structured signal from unknown jamming attacks. To identify jamming or unauthorised wireless access, in [17], the authors proposed a channel state information based approach. With a quaternary hypotheses test, the detection framework focuses on distinguishing between legitimate and illegitimate transmissions. In a generalised multivariate analysis of variance signal model with structured interference, the maximum invariant statistic was used to create appropriate detectors with a constant false alarm rate [18]. In [19], the authors used the second-order statistics of the received signals to generate five sub-optimal fusion rules for detecting attacks.

Although many methods have been proposed for jamming detection [11]–[19], very few methods were designed considering the jamming attacks for mission-critical mMTC applications, and existing methods are not suitable for UAJ detection. This is mainly because of the following reasons:
 
1) Most existing statistic features are not robust to discriminate between UAJ and other causes of fluctuations induced by the legitimate communication itself, such as the energy characteristic [15], subspace dimension [16], and channel state information [17]. The maximum invariant statistics [18] and the second-order statistics [19] need to apply a complex multi-antenna mechanism, which is not affordable for low-cost MTDs. In addition, statistical features of upper layer [11]-[14] will cause an intolerable delay.

2) Existing detection methods are mostly implemented in a batch manner, namely, all sample observations are collected before making a decision. However, our goal is to detect whether the newly arriving samples in current frame is anomalous due to UAJ. The detection is not only one time and end, namely, not all sample observations for the whole transmission process are collected before making a decision. The false alarms of the batch detection are prone to be raised for the frame based detection with a small number of time slots, which is not suitable for the considered detection task of this paper.

3) Most of the related works rely on the prior knowledge of the attacker to select a decision threshold for distinguishing the jamming attack from the normal state, which is unrealistic. Because the adversary will not cooperate with the legitimate system, it is not easy to obtain these statistics.

\subsection{Motivations and Contributions}

In order to design a timely and reliable method for UAJ detection, in this paper, we propose to exploit the joint space-time sparsity (JSTS) based feature. Our JSTS detection method is motivated by the fact that the original joint space-time sparsity will be destroyed if UAJ exists. Specifically, in the typical mMTC scenarios, the common features of the received uplink access signals are the space sparsity and the time sparsity. The space sparsity refers to the fact that only a small subset of devices are active in order to save energy most of the time in mMTC scenarios, and the time sparsity is usually caused by the fact that the traffic pattern for each device is sporadic. These two intrinsic characteristics of mMTC are the key points in the realization of joint active device and data detection. As soon as UAJ occurs, the original joint space-time sparsity will be destroyed. This is because that UAJ attackers will send numerous access jamming signals in a short period of time in order to disrupt the device activity and data detection, which can cause the increase of the number of access devices and lead to the destruction of sporadic device activity. In other words, the JSTS-based feature can well characterize the transmission pattern of mMTC, leading to better attack detection performance. 

In the proposed JSTS-based method, we first extract the joint space-time sparsity by solving a joint space-time sparsity constrained maximum likelihood factor analysis problem. When UAJ exists, the BS can detect the attack through checking the abrupt change of the JSTS-based feature between two adjacent frames, referred to as sequential change frame detection. The main contributions of the proposed JSTS-based detection method lie in the following three aspects:

1) To extract the joint space-time sparsity, we solve a challenging joint space-time sparsity constrained factor analysis problem. Factor analysis is a popular multivariate dimensionality reduction method for determining the structure of correlations between a set of observed random variables. Specifically, we first present a reformulation of this challenging problem to an equivalent optimization problem without explicit space sparsity constraint, and then develop an effective approach based on difference of convex optimization to extract the JSTS-based feature.

2) Different from existing detection methods under batch manner, we perform the JSTS-based detection in a sequential manner, which takes one observation at a time without having to re-explore all previously available observations [20]. Under sequential detection, if no change of the JSTS-based feature is detected, then the detector moves to the next frame instant, which can detect the attack as quickly as possible.

3) Since the proposed JSTS-based method only cares the abrupt change in the JSTS-based feature on each frame, we do not need to know the accurate prior information of the attacker. Based on the joint space-time
sparsity constrained factor analysis, the proposed JSTS-based method can be efficiently carried out by a solvable difference of convex functions, and can detect UAJ in a real-time manner by using sequential change frame detection. 

\begin{figure*}[!t]
 	\centering
 	\includegraphics[width=6.8 in, height=2.3 in]{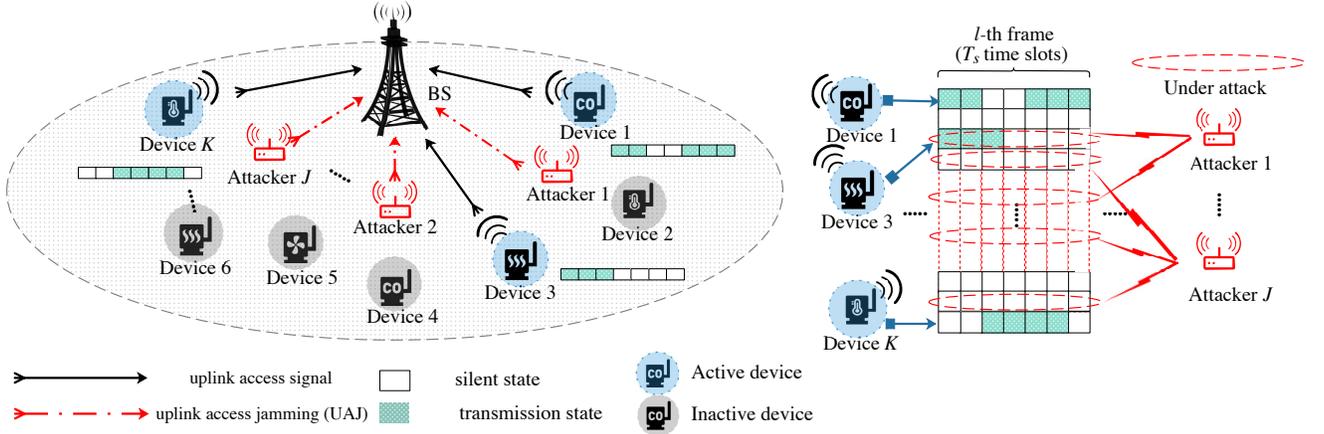}
 	\caption{ System model.}
 	\hrulefill
 \end{figure*}

\emph{Organization:} In Section II, we present the system model with UAJ attackers. In Section III, we first describe the problem of UAJ detection, and then introduce the basic detection principle of the proposed JSTS-based method. We present the complete detection framework for the JSTS-based method and give the corresponding computation complexity analysis in Section IV. Finally, we evaluate the performance of the JSTS-based method by some numerical results in Section V, and conclude the work in Section VI.

\emph{Notations:} $\pmb{I}_M$ denotes a $M\times M$ identity matrix. Diagonal matrix is denoted by $\rm{diag}(\cdot)$. The column-ordered vectorization of matrix $\bm X$ is ${\bm x} = {\rm vec}(\bm X)$. $||\cdot||_0$ denotes the $l_0$ norm. ${\bm \Gamma} / {\bm \Omega}$ represents the set composed of elements in ${\bm \Gamma} $ but not in set ${\bm \Omega}$. The number of elements in the set ${\bm \Gamma}$ is calculated by $|{\bm \Gamma}|$. $\Re(\cdot)$ and $\Im(\cdot)$ denote real and imaginary parts of a complex number. $\mathbb{C}^{N\times M}$ and $\mathbb{R}^{N\times M}$ denote the spaces of all $N\times M$ matrices with complex-valued and real-valued elements, respectively. $\mathbb{CN}(\pmb{\mu},\pmb{\bm \Sigma} )$ and $\mathbb{N}(\pmb{\mu},\pmb{\bm \Sigma} )$ denote the distributions of complex and real Gaussian random vectors, respectively, with mean $\pmb{\mu}$ and covariance matrix $\pmb{\bm \Sigma}$.

\section{System Model}

\subsection{System Model}

We consider the uplink of a mission-critical mMTC communication scenario, where $Q$ MTDs activated from $K$ potential devices simultaneously transmit uplink access signals to the BS in the presence of $J$ attackers, who aim to disturb the uplink access procedure by jamming, i.e., UAJ. For simplicity, we assume that the BS and the MTDs as well as the attackers have a single antenna. Note that our method can also be applied to the cases when the BS and the MTDs are equipped with multiple antennas [21]. This is because the feature extraction of the joint space-time sparsity can be divided into several parallel sub-problems at each receive antenna. As illustrated in Fig. 1, the typical uplink transmission in mMTC communications have two distinctive characteristics: 1) Space sparsity: only a few MTDs are active, although the number of potential devices may be very large; 2) Time sparsity: the transmitted symbols of the MTDs are typically sporadic in successive time slots within a frame.

In the normal case with grant-free random access scheme, each active MTD can directly transmit its access requests without being scheduled, and is preassigned with a unique indicator sequence as the identification of the uplink access signal to realize joint device activity and data detection. We consider a consecutive time slots dynamic model, the superscript ${\left\{ \cdot \right\}^{( {l,t} )}}$ denotes the $t$th time slot ($t = 1,2, \cdots ,T_s $) in the $l$th frame ($l = 1,2, \cdots ,L $). We use $d_k^{( {l,t} )}$ to denote the transmitted symbol of the $k$th MTD, which are randomly generated from the Quadrature Phase Shift Keying (QPSK) symbol set [22]. We denote $a_k^{( {l,t} )}$ as the activity indicator which is set to 1 or 0 if the $k$th MTD is active or inactive. Due to the time sparsity of device activity, most MTDs are idle and do not transmit any symbols, which indicates that $\bm{a}_{}^{( {l,t} )} = {\left[ {a_1^{( {l,t} )},a_2^{( {l,t} )}, \ldots ,a_K^{( {l,t} )}} \right]^T}$ is a sparse vector and ${\left\| {\bm{a}_{}^{( {l,t} )}} \right\|_0} \ll K$ [23], [24].

To characterize the time-variation of the active device indicator $\bm{a}_{}^{( {l,t} )}$, we consider a probabilistic model to characterize the time-variation of the active MTD indicator, where MTDs are supposed to form independent Markov chains by a couple of transition probabilities $p_{}^{(01)}\triangleq \{ {a_k^{( {l,t} )} = 0| {a_k^{( {l,t - 1} )} = 1}} \}$. Under this condition, the Markov chain of each MTD is under a steady state with $\Pr \{ {a_k^{( {l,t} )} = 1} \} = \mu _k$ that indicates the activation probability of each MTD and can be specified by the transition probability $p_{}^{(01)}$ and the active device ratio $\mu _k$. Moreover, we consider the transition probabilities are independent with $k$, denoted by $p_{}^{(01)} = \rho ( {1 - \mu } )$, where $\rho$ denotes a scale factor that controls the specific value of transition probability and $\mu $ denotes the active device ratio. This model has shown its efficiency in the similar application [25]. Denote the spreading sequences allocated to the MTDs as $\left\{ {\bm{s}_k^{}} \right\}_{k = 1}^K$, the transmitted symbol $d_k^{( {l,t} )}$ is modulated by a device-specific spreading sequence $\bm{s}_k^{} = [ s_{1,k}^{},s_{2,k}^{},$ $\ldots, s_{N,k}]^T$ of length $N$. The spreading sequence of each MTD is designed as a Gaussian vector whose components are realized from independent and identical distributed (i.i.d.) Gaussian sources $\mathbb{N}(0,1)$. Note that the proposed method can also be applied to the cases where the spreading sequences are generated using Hadamard matrices [26], since the joint space-time sparsity based feature, the detection method and the conclusion for the considered problem in this paper are not affected. We use the parameter ${\eta ^{( l )}}$ to indicate the correlation between the active device set in the adjacent time slots, e.g., ${\eta ^{( l )}} \triangleq \left| {{{\bm \Gamma} ^{( {l,t - 1} )}} \cap {{\bm \Gamma} ^{( {l,t} )}}} \right|/\left| {{{\bm \Gamma} ^{( {l,t} )}}} \right|$, where ${{\bm \Gamma} ^{( {l,t} )}} \triangleq \left\{ {k:d_k^{( {l,t} )} \ne 0,1 \le k \le K,\forall k} \right\}$, and $\left| {{{\bm \Gamma} ^{( {l,t} )}}} \right|$ is the cardinality of active device set in the $t$th time slot. A larger ${\eta ^{( l )}}$ means a stronger temporal correlation. Then, the received uplink access signals from all active MTDs at the BS, denoted by $\bm{y}_{}^{( {l,t} )}$, is given by
\begin{align}
 \bm{y}_{}^{( {l,t} )} & = \sum\limits_{k = 1}^K {a_k^{( {l,t} )}\sqrt {{P_k}{\beta _k}} }\bm{s}_k^{}h_k^{(l,t)}d_k^{( {l,t} )} + \bm{w}  \cr 
  & = \bm{S}{\bm{d}^{( {l,t} )}} +\bm{w},
\end{align}
where ${P_k}$ is the transmit power of the $k$th MTD, and ${\beta _k}$ is the distance-based path losses from the $k$th MTD to the BS. $h_k^{(l,t)}\sim\mathbb{C}\mathbb{N}( 0,1  )$ indicates the corresponding fading channel between the BS and the $k$th MTD. We assume that all the channel coefficients are i.i.d. over different time slots. $\bm{w}$ is the additive white Gaussian  noise (AWGN) vector with each element being distributed as $\mathbb{C}\mathbb{N}( 0, \sigma_w^2  )$.  We further define $ \bm{S} \triangleq {[ {\bm{s}_1^{}; \cdots ;\bm{s}_K^{}} ]}$ and ${ \bm{d}^{( {l,t} )}} \triangleq {[ {a_1^{( {l,t} )}\sqrt {{P_1}{\beta _1}}h_1^{( {l,t} )}d_1^{( {l,t} )}, \ldots ,a_K^{( {l,t} )}\sqrt {{P_K}{\beta _K}}h_K^{( {l,t} )}d_K^{( {l,t} )}} ]^T}$.  

In mission-critical mMTC scenarios with grant-free random access scheme, the most important features of the received uplink access signals $\bm{y}_{}^{( {l,t} )} $ is the joint space-time sparsity. On one hand, in order to save energy, only a small portion of potential devices are active in most cases, referred to as the space sparsity, i.e., $\bm{a}_{}^{( {l,t} )}$ is a sparse vector. On the other hand, the transmitted signals $\bm d^{( {l,t} )}$ is sparse since $\bm{d}_{}^{( {l,t} )}$ is typically sporadic in successive time slots within a frame [26], [27], referred to as the time sparsity. Note that the device activity is a critical issue for mission-critical mMTC applications with grant-free random access scheme [27]. Since the BS does not have the information of device activity, in the absence of scheduling, the multi-device detection is needed before the data detection in order to distinguish active MTDs from other inactive MTDs.

By exploiting the joint space-time sparsity, compressive sensing (CS) technology [28] is therefore promising to realize joint active device and data detection corresponding to a sparse signal recovery problem, e.g., the approximate message passing algorithm [6] based on the CS, especially the temporal correlation can be utilized to improve compressive detection performance for reconstruct these sparse signals.

\subsection{Attack Model}

Due to the weak security protection capabilities of low-cost low-power MTDs, mission-critical mMTC application with the grant-free random access mechanism is very vulnerable to UAJ. Note that in the grant-free mMTC, since the spreading sequence is also used for MTD identification, it is usually fixed for each MTD and cannot be randomized for each transmission. UAJ attackers can easily trick the BS under the cover of their legitimate identities by eavesdropping the legitimate MTDs’ spreading sequences [7]. For instance, since the spreading sequence resource pool and the time-frequency plane for grant-free based uplink access are publicly known, the attackers could easily obtain these information. Furthermore, some UAJ attackers may even hijack legitimate MTDs, which means that they can establish legitimate connections with the BS and acquire the information of uplink transmission resources. Hence, in this paper, we mainly consider that the attackers know the spread matrix, which can be treated as the worst case. If the detection performance of the proposed method mets the needs in service even for the worst case, it means that the proposed method has application value in practice. 

We define $\bm{\Omega }$ as the set of a spreading sequence resource pool. MTDs and the attackers can randomly select the spreading sequences from $\bm{\Omega }$. Then, we use ${\bm{\Omega } _Q} \subseteq \bm{\Omega }  $ and ${\bm{\Omega } _J} \subseteq \bm{\Omega }  $ to denote the set of spreading sequences selected by the active MTDs ($\bm{s}_q^{} \in {\bm{\Omega } _Q}, q \in \left\{ {k\left| {a_k^{( {l,t} )} = 1} \right.} \right\}$) and the attackers ($\bm{s}_{A,j}^{} \in {\bm{\Omega  }_J}$) during the uplink access, respectively, where $\bm{s}_{A,j}^{}$ denotes the spreading sequences selected by the $j$th attacker. Given ${\bm{\Omega } _J} \triangleq  {\bm{\Omega } _{J_1}} \cup {\bm{\Omega } _{J_2}}$, a part of UAJ attackers who select the spreading sequences from ${\bm{\Omega  }_{J_1}}$ mainly influence the data detection when ${\bm{\Omega } _{J_1}}$ is a subset of ${\bm{\Omega } _Q}$, and another part of UAJ attackers who select the spreading sequences from ${\bm{\Omega  }_{J_2}}$ predominantly affect the device activity detection when ${\bm{\Omega } _Q} \cap {\bm{\Omega } _{J_2}}= 0$. In the presence of UAJ, the received signal $\tilde {\bm{y}}_{}^{( {l,t} )}$ at the BS can be modeled as 
\begin{align}
\tilde {\bm{y}}_{}^{( {l,t} )} &= \bm{S}{\bm{d}^{( {l,t} )}}  + \sum\limits_{j = 1}^J {\sqrt {{P_{A,j}}{\beta _{A,j}}} } \bm{s}_{A,j}^{}g_j^{(l,t)}u_j^{( {l,t} )} + \bm{w}  \cr 
  & =\bm{S}( {{\bm{d}^{( {l,t} )}} + {\bm{u}^{( {l,t} )}}} ) + \bm{w },
\end{align}
where ${P_{A,j}}$ is the transmit power of the $j$th attacker, and ${\beta _{A,j}}$ is the distance-based path losses from the $j$th attacker to the BS. ${g}_j^{( {l,t} )}$ indicates the corresponding fading channel between the BS and the $j$th attacker, which is distributed as $\mathbb{C}\mathbb{N}( 0,1  )$. $u_j^{( {l,t} )}$ denotes the transmitted symbol of the $j$th attacker, which is generated from i.i.d. Gaussian random variables with zero mean and unit variance. Define the spreading sequences selected by the $j$th attacker as $\bm{s}_{A,j}^{} \triangleq \sum\limits_{k = 1}^K {\theta _{j,k}^{}\bm{s}_k^{}} $, and $\theta _{j,k}$ is the adjustment allocation coefficient allocated for attacking the access of the $k$th MTD. When $\theta _{j,k}=0$,  the $k$th MTD is not affected by the $j$th attacker, and vice versa. 

\begin{figure*}[!t]
 	\centering
 	\includegraphics[width=4 in]{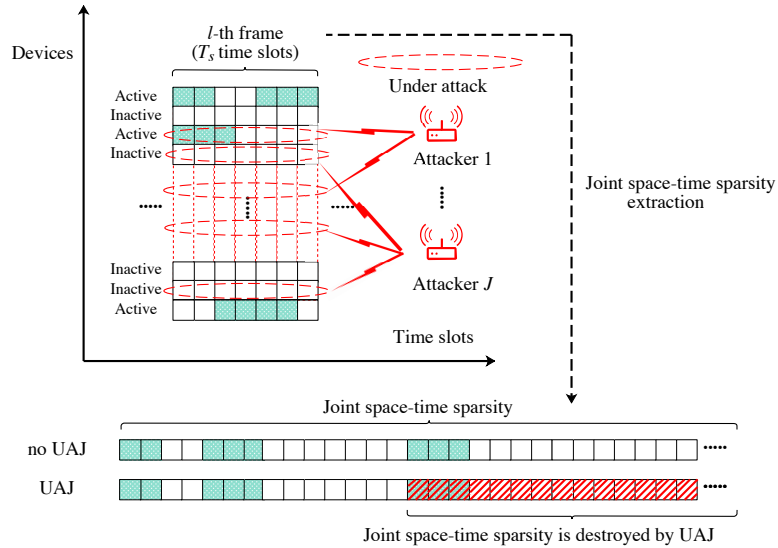}
 	\caption{Frame structure of uplink access signals and the effect of UAJ.}
 	\hrulefill
\end{figure*}

Because MTDs occur sporadically, it is difficult for the attacker to obtain the accurate priori information of the underlying MTD activity characteristics. Thus, in this paper, the allocation coefficient $\theta _{j,k}$ is realized from i.i.d. uniform distribution $\mathbb{U}(0,1)$. Note that optimizing the allocation coefficient $\theta _{j,k}$ to minimize the detection probability from the perspective of the attacker is beyond the scope of this paper. We also have studied a similar problem in [29]. This interesting issue requires further investigation in our future research. We further define $\bar P_{A,j}\triangleq {\sqrt {{P_{A,j}}{\beta _{A,j}}} }$, and ${\bm{u}^{( {l,t} )}} \triangleq {\left[ {\sum\limits_{j = 1}^J {\bar P_{A,j} } {\theta _{j,1}^{}g_j^{( {l,t} )}u_j^{( {l,t} )}}, \ldots ,\sum\limits_{j = 1}^J {\bar P_{A,j} } {\theta _{j,K}^{}g_j^{( {l,t} )} u_j^{( {l,t} )}} } \right]^T}$ which will affect the correctness of device activity and data detection. In the following sections, our proposed JSTS-based method and the corresponding performance analysis are closely interrelated to this attack model.

\emph{Remark:} We would like to point out that in this paper, we do not consider the case where the attackers send the spreading sequences $\bm{s}_{A,j}$ in an intermittent manner. This is mainly because such kind of attack strategy suffers from two major weaknesses. On one hand, it is hard for the attackers to obtain the accurate priori information of the underlying MTD activity characteristics since MTDs occur sporadically. Thus, it can be extremely difficult to perform an efficient attack if the attackers send spreading sequences in an intermittent manner. On the other hand, if the attack carried out blindly regardless of the MTDs’ activity, the attack effect is limited and quite time consuming due to the extremely low collision probability between the legitimate MTDs and the attackers. Furthermore, if the number of the attackers increases in order to improve the attack performance, our method can also handle this case since the space sparsity has changed dramatically. 

\section{UAJ Detection via JSTS-based method}

In this section, we first construct the detection problem of UAJ and introduce detection principle of the proposed JSTS-based method. Then, details are given for the joint space-time sparsity constrained factor analysis problem for the JSTS-based feature extraction.

\subsection{Problem Statement}

For the uplink access in mission-critical mMTC application with some UAJ attackers, the problem of UAJ detection becomes how to distinguish the superimposed signal ${\bm{y}}_{}^{( {l,t} )}$ in (1) from $\tilde {\bm{y}}_{}^{( {l,t} )}$ in (2). Denote ${\bm{o}^{( {l,t} )}}$ as the observation signals received by the BS, we have that
\begin{align}
{\bm{o}^{( {l,t} )}}=\left\{
\begin{aligned}
&{\bm{y}_{}^{( {l,t} )} =\bm{S}{\bm{d}^{( {l,t} )}} +\bm{w}}, && \rm {no\ UAJ},\\
&{\tilde {\bm{y}}_{}^{( {l,t} )} =\bm{S}( {{\bm{d}^{( {l,t} )}} + {\bm{u}^{( {l,t} )}}} ) + \bm{w }}, && \rm UAJ.
\end{aligned}\right.
\end{align}

Factor analysis is a classical multivariate dimensionality reduction technique, and is an extension of principal component analysis. Factor analysis is a popular method for determining the structure of correlations between a set of observed random variables [30], [31]. Considering the factor analysis adopted in the proposed JSTS-based method is implemented in the real-valued system, the complex detection model in (3) is expanded as a real one $\bm{o}_r^{( {l,t} )} \in {{\mathbb R}^{2N}}$, which can be represented as
\begin{align}
\label{DetectionModel}
{\bm{o}_r^{( {l,t} )}}=\left\{
\begin{aligned}
&{\bm{y}_{r}^{( {l,t} )} =\bm{S}_r{\bm{d}_r^{( {l,t} )}} +\bm{w}_r}, && \rm {no\ UAJ},\\
&{\tilde {\bm{y}}_{r}^{( {l,t} )} =\bm{S}_r{\tilde {\bm{d}}_r^{( {l,t} )}}+ \bm{w }_r}, && \rm UAJ,
\end{aligned}\right.
\end{align}
where 
$\begin{matrix}
\bm{o}_r^{( {l,t} )} = \left[ {\matrix
   {\Re ( {{\bm{o}^{( {l,t} )}}} )}  \\ 
   {\Im ( {{\bm{o}^{( {l,t} )}}} )}  \\ 

 \endmatrix } \right]
\end{matrix}$, 
$\begin{matrix}
\bm{S}_r^{} = \left[ {\matrix
   {\Re ( \bm{S} )} & { - \Im ( \bm{S} )}  \\ 
   {\Im ( \bm{S} )} & {\Re( \bm{S} )}  \\ 

 \endmatrix } \right]
\end{matrix}$,
$\begin{matrix}
\bm{d}_r^{( {l,t} )} = \left[ {\matrix
   {\Re( {{\bm{d}^{( {l,t} )}}} )}  \\ 
   {\Im ( {{\bm{d}^{( {l,t} )}}} )}  \\ 

 \endmatrix } \right]
\end{matrix}$,
$\tilde {\bm{d}}_r^{( {l,t} )} =$ 
$\begin{matrix}
\left[ {\matrix
   {\Re( {{{\tilde {\bm{d}}}^{( {l,t} )}}} )}  \\ 
   {\Im ( {{{\tilde {\bm{d}}}^{( {l,t} )}}} )}  \\ 

 \endmatrix } \right]
\end{matrix}$,
$\tilde {\bm{d}}^{( {l,t} )} = {\bm{d}^{( {l,t} )}} + {\bm{u}^{( {l,t} )}}$,
$\begin{matrix}
\bm{w}_r^{} = \left[ {\matrix
   {\Re( \bm{w} )}  \\ 
   {\Im ( \bm{w} )}  \\ 

 \endmatrix } \right]
\end{matrix}$.
To effectively design UAJ detection method, we first collect the observation signals received by the BS in consecutive $T_s$ time slots in the $l$th frame, and transfer this observation into an equivalent vector as $\bm{o}_r^{( l )} = {\rm {vec}} ( {{{\left[ {\bm{o}_r^{( {l,1} )},\bm{o}_r^{( {l,2} )}, \cdots ,\bm{o}_r^{( {l,{T_s}} )}} \right]}^T}} )$. Then we extract the joint space-time sparsity for UAJ detection, as illustrated in Fig. 2.

Factor analysis [30], [31] usually models the traffic monitoring data of a time slot as a vector and use a traffic matrix to record the traffic monitoring data of a period. As normal traffic data generally exhibit strong spatio-temporal correlations, factor analysis usually separates the observed traffic data into two parts, a low-rank normal data matrix and a sparse outlier data caused by the noise or wireless channel. Accordingly, in this paper, $\bm{o}_r^{( l )}$ can be decomposed as two parts $\bm{o}_{r}^{( l )} = \bm{v}^{( l )} + \bm{\iota}^{( l )}$. The first part $\bm{v}^{( l )}$ is a vector from a low-rank subspace ${\bm{V}^{( l )}}$, and the second part $\bm{\iota}^{( l )}$ is a sparse error with support size ${\tau ^{( l )}}$. Here the number of nonzero items in $\bm{\iota}^{( l )}$ is represented by ${\tau ^{( l )}} = \sum\limits_{i = 1}^{\phi} \delta_D ({\bm{\iota}_i^{( l )} \ne 0})$, where $\delta_D(\cdot)$ is the Dirichlet function and $\phi=2N{T_s}$. The underlying subspace ${\bm{V}^{( l )}}$ might or might not change with time $t$. When ${\bm{V}^{( l )}}$ does not change over time, we say the data are generated from a stable subspace. Otherwise, the data are generated from a changing subspace. Hence, factor analysis can be used to anomaly detection by tracking the abruptly changing subspaces caused by abnormal data. This motivates us to exploit the joint space-time sparsity through the sparsity constrained factor analysis to detect access jamming in mission-critical mMTC. Based on the above analysis, in the next subsection, we introduce detection principle of our proposed JSTS-based method.

\subsection{Detection Principle}

Our JSTS-based detection method relies on the following facts: 1) Absence of UAJ: when no UAJ is present, according to the received uplink access signals $\bm{o}_{}^{( {l,t} )} = \bm{y}_{}^{( {l,t} )}$ in (1), ${\rm {Cov}}( {\bm{o}_r^{( l )}} ) \triangleq \bm{ \Sigma}^{(l)} $ has the joint space-time sparsity since $\bm{a}_{}^{( {l,t} )}$ is a sparse vector and ${\left\| {\bm{a}_{}^{( {l,t} )}} \right\|_0} \ll K$, and the transmitted signals $\bm{d}_{}^{( {l,t} )}$ is typically sporadic in successive time slots within a frame. Base on the joint space-time sparsity, the joint active device and data detection can be processed effectively; 2) Presence of UAJ: in the presence of UAJ, refer to the superimposed signals $\bm{o}_{}^{( {l,t} )} = \tilde {\bm{y}}_{}^{( {l,t} )}$ in (2), we notice that the joint space-time sparsity will be destroyed. This is due to the fact that the UAJ attackers aim to interfere the results of device activity and data detection by jamming, which will inevitably result in the destruction of the joint space-time sparsity. 

Thus, the detection principle of the proposed JSTS-based method can be summarized as follows. If UAJ does not exist, the current joint space-time sparsity in the $l$th frame ${\bm{o}_r^{( l )}}$ and the recorded characteristic in $(l-1)$th frame ${\bm{o}_r^{( l-1 )}}$ should be similar. Conversely, when UAJ launches, the joint space-time sparsity will be destroyed. Thus, the abrupt change of the joint space-time sparsity between the newly arriving samples in the current $l$th frame and the monitoring samples in the previous $(l-1)$th frame can be used to detect whether UAJ happens in the current $l$th frame. 

\subsection{Joint Space-time Sparsity Constrained Factor Analysis}

In order to extract the JSTS-based feature, it is necessary to go into the following details of joint space-time sparsity constrained factor analysis before describing the complete JSTS-based framework, including \emph{problem construction} and \emph{problem reformulation}.

\emph{Problem construction:} The JSTS-based feature extraction leads to a challenging joint space-time sparsity constrained factor analysis problem. The maximum likelihood principle, which aims to maximise the Gaussian likelihood, is commonly used in factor analysis estimation. For more details about factor analysis, please refer to [30], [31] and references therein. The maximum likelihood (ML) based approach [32] is one of the most widely used factor analysis estimate methods. The purpose of the ML-based method is to minimize the negative log-likelihood with regard to signal covariance matrix $\bm{ \Sigma}^{(l)}$. The expectation maximization (EM) [33] is another popular method for factor analysis. To ensure a fair comparison, the joint space-time sparsity constraint is also considered in the ML-based and the EM-based methods. Specifically, we reformulate the joint space-time sparsity constrained ML-based factor analysis task as a nonlinear nonsmooth semidefinite optimization problem, which can be solved by using the nuclear norm relaxations of the joint space-time sparsity constraint [32]. In addition, the joint space-time sparsity constrained EM-based factor analysis problem can be tackled by a spatial branch and bound method for minimizing the squared Frobenius norm as in [33]. 

Note that this paper is aimed at timely detection of UAJ, and our goal is to detect whether the newly arriving samples $\left\{ {{\bm{o}^{( {l,t} )}}} \right\}_{t = 1}^{{T_s}}$ in the current $l$th frame is anomalous due to UAJ. According to the typical mMTC scenario in the 5G new radio specification [5], the number of the consecutive time slots $T_s$ in each frame is limited. However, the ML-based and EM-based methods are all based on the empirical covariance matrix via the maximum likelihood principle and are far from satisfactory for the frame based detection with a small number of time slots in this paper. Note that in addition to a slow convergence speed, the ML-based and EM-based methods seem to get stuck in suboptimal local solutions [32], [33]. In view of these issues, we need to propose a new computational framework for solving the following joint space-time sparsity constrained factor analysis problem to achieve the timely detection goal in this paper (we drop the superscript of index $l$ for conciseness)
\begin{align} 
\nonumber
{(P1):}~~~\hbox {min}~~~&{{\mathcal {D}}}(\bm{\Sigma })\triangleq - \log \det (\bm{\Sigma }^{-1}) + {{\rm {tr}}}({\bm{\Sigma }}^{-1}{{\bm {R}}^{}}), \\ \nonumber \rm {s. t.}~~~&{\bm{\Sigma }} = \bm{{\bm{P}^{}} } + \bm{\Xi }, {\rm {rank}}(\bm{\Xi }) \le r, \\& \bm{{\bm{P}^{}} }={{\rm {diag}}}({{p}^{}} _1, \ldots , {{p}^{}}_\phi) \succeq \epsilon \bm {I} \nonumber, 
\end{align}
where ${{\bm {R}}^{}}$ is the sample covariance matrix of $\bm{o}_r$, $\bm{\Xi }\triangleq {\bm{V}^{}} {\bm{V}^{}}^T$, $r$ is the space sparsity constraint, $\epsilon$ is the time sparsity indicator and $\bm {I}$ is the identity matrix. $\bm{P}$ and $\bm{\Xi}$ are the optimization variables. Except for the additional space-time sparsity constraints, the objective function in problem (P1) is not a concave function ${\mathcal {D}}(\bm{\Sigma })$ and equality constraint of $\bm{\Sigma }$ is nonconvex, thus, problem (P1) is a non-convex optimization problem and difficult to solve. In the following, we introduce how to reformulate problem (P1) to a solvable difference of convex functions.

\emph{Problem reformulation:} We first present a reformulation of problem (P1) to an equivalent optimization problem (P2) in which there is no explicit space sparsity constraint by the following proposition, and then effective optimization approaches based on difference of convex optimization [34], [35], can be used to solve the problem (P2).

\emph{Proposition 1: Define ${{\bm {R}}^{}}^{\prime} \triangleq \bm{{\bm{P}^{}} }^{-\frac{1}{2}} {{\bm {R}}^{}} \bm{{\bm{P}^{}} }^{-\frac{1}{2} }$, and let $\lambda _1^{\prime} \ge \lambda _2^{\prime} \ge \cdots \ge \lambda _\phi^{\prime}$ denote the eigenvalues of ${{\bm {R}}^{}}^{\prime}$, then problem (P1) is equivalent to the following problem (P2) and the optimization variables are $\bm{P}$ and $\bm {R}^{\prime}$
}
\begin{align} 
\nonumber
{(P2):}~~~\hbox {min}\;\;&\displaystyle \left\{ \log \det (\bm{{\bm{P}^{}} }) + {\rm{tr}} ( {{\bm {R}}^{}}^{\prime} ) + e_r(\lambda _i^{\prime}) \right\}, \\\nonumber \rm{s. t.}\;\;\;&\bm{{\bm{P}^{}} }={\rm{diag}}({{p}^{}} _1, \ldots , {{p}^{}} _\phi) \succeq \epsilon \bm {I} ,
\end{align}
where $e_r(\lambda _i^{\prime})\triangleq \sum ^{r}_{i=1} ( \log (\max \{ 1,\lambda _i^{\prime}\})-\max \{1,\lambda _i^{\prime}\} +1 )$.

\emph{Proof:} For a fixed $\bm{P}$, we first minimize problem (P1) with regard to $\bm{V}$. Based on a straightforward application of the S. W. formula [36], we have that 
\begin{align} 
\bm{\Sigma }^{-1}= & {} \bm{{\bm{P}^{}} }^{-1} - \bm{{\bm{P}^{}} }^{-1} {\bm{V}^{}} \bm {W_{P,V}},
\end{align}
where $\bm {W_{P,V}}\triangleq ( \bm {I} + {\bm{V}^{}}^T \bm{{\bm{P}^{}} }^{-1} {\bm{V}^{}}) ^{-1} {\bm{V}^{}}^T \bm{{\bm{P}^{}} }^{-1}$. By substituting ${\bm{V}^{}}^{\prime} \triangleq \bm{{\bm{P}^{}} }^{-\frac{1}{2}} {\bm{V}^{}}$, we have ${\rm {tr}}(\bm{\Sigma }^{-1}{{\bm {R}}^{}})={\rm {tr}}({{\bm {R}}^{}}^{\prime}) - {\rm {tr}}( ({\bm{V}^{}}^{\prime})^T {{\bm {R}}^{}}^{\prime}{\bm{V}^{}}^{\prime} ( \bm {I} + ({\bm{V}^{}}^{\prime})^T {\bm{V}^{}}^{\prime} ) ^{-1} )$. It is worth noting that $-\log \det (\bm{\Sigma }^{-1}) = {} \log \det (\bm{{\bm{P}^{}} }) + \log \det ( \bm {I} + ({{\bm{V}^{}}^{\prime}})^T {\bm{V}^{}}^{\prime} )$. By combining ${\rm {tr}}(\bm{\Sigma }^{-1}{{\bm {R}}^{}})$ and $-\log \det (\bm{\Sigma }^{-1})$, problem (P1) can be rewritten as 
\begin{align} 
\hbox {min}~~&\log \det (\bm{{\bm{P}^{}} }) + f_r({\bm{R}}^{\prime},{\bm{V}}^{\prime} )\nonumber,\\ \rm {s. t.}~~&\bm{{\bm{P}^{}} }={\rm {diag}}({{p}^{}} _1, \ldots , {{p}^{}} _\phi) \succeq \epsilon \bm {I} \nonumber,
\end{align}
where $f_r({\bm{R}}^{\prime},{\bm{V}}^{\prime} )\triangleq \log \det ( \bm {I} + ({{\bm{V}^{}}^{\prime}})^T {\bm{V}^{}}^{\prime} ) + {\rm {tr}} ( {{\bm {R}}^{}}^{\prime} ) - {\rm {tr}}( ({{\bm{V}^{}}^{\prime}})^T {{{\bm {R}}^{}}^{\prime}} {\bm{V}^{}}^{\prime} ( \bm {I} + ({{\bm{V}^{}}^{\prime}})^T {\bm{V}^{}}^{\prime}) ^{-1})$. We use $h(\bm{{\bm{P}^{}} },{\bm{V}^{}})\triangleq \log \det ({\bm{V}^{}}{\bm{V}^{}}^T + \bm{{\bm{P}^{}} }) + {\rm {tr}}(({\bm{V}^{}}{\bm{V}^{}}^T + \bm{{\bm{P}^{}} })^{-1} {{\bm {R}}^{}})$ to denote objective in problem (P1). The partial derivative of $h(\bm{{\bm{P}^{}} },{\bm{V}^{}})$ with respect to $\bm{V}$ can be obtained by $\frac{\partial h(\bm{{\bm{P}^{}} },{\bm{V}^{}})}{\partial {\bm{V}^{}}} = 2\bm{\Sigma }^{-1}( {\bm{\Sigma }} - {{\bm {R}}^{}} ) \bm{\Sigma }^{-1}{\bm{V}^{}}$. Note that $\partial h(\bm{{\bm{P}^{}} },{\bm{V}^{}})/\partial {\bm{V}^{}} =0$ if and only if ${\bm{V}^{}} = {{\bm {R}}^{}}( \bm{{\bm{P}^{}} }+ {\bm{V}^{}} {\bm{V}^{}}^T )^{-1}{\bm{V}^{}}$. By using algebraic manipulations with help of (5), this condition can be reformulated as ${{\bm {R}}^{}}^{\prime}{\bm{V}^{}}^{\prime} = {\bm{V}^{}}^{\prime}( \bm {I} + {({\bm{V}^{}})^{\prime}}^T {\bm{V}^{}}^{\prime} )$.

Because we chose pairwise orthogonal or zero vectors for the columns of ${\bm{V}^{}}^{\prime}$, $\bm {I} + ({{\bm{V}^{}}^{{\prime} }})^T {\bm{V}^{}}^{\prime}$ is a diagonal matrix with every diagonal element bigger than or equal to one. As a result, ${{\bm {R}}^{}}^{\prime}{\bm{V}^{}}^{\prime}$ is a collection of eigenvector equations for the matrix ${{\bm {R}}^{}}^{\prime}$. Let ${{\bm{\zeta}}_i}, i \in [r]$ be the columns of ${\bm{V}^{}}^{\prime}$, and define $g({\bm{V}^{}}^{\prime}) \triangleq \sum _{i=1}^{r} ( \log (1 + {{\bm{\zeta}}_i}^T {{\bm{\zeta}}_i}) - \frac{{{\bm{\zeta}}_i}^T {{\bm {R}}^{}}^{\prime} {{\bm{\zeta}}_i}}{1 + {{\bm{\zeta}}_i}^T {{\bm{\zeta}}_i}} )$. By combining ${{\bm {R}}^{}}^{\prime}{\bm{V}^{}}^{\prime}$, because ${{{\bm{\zeta}}_i}}$ are pairwise orthogonal or zero vectors, we have that ${{\bm {R}}^{}}^{\prime}{{\bm{\zeta}}_i}={\kappa_i} {{\bm{\zeta}}_i} $ and ${\kappa_i}=1+{{\bm{\zeta}}_i}^T {{\bm{\zeta}}_i}$, $i \in [r]$. Note that either ${\kappa_i} =1$ with ${{\bm{\zeta}}_i} = 0$ or ${\kappa_i} > 1$ and ${\kappa_i}$ equals some eigenvalue of ${{\bm {R}}^{}}^{\prime}$ with eigenvector ${{\bm{\zeta}}_i}$, therefore $g({\bm{V}^{}}^{\prime})=\sum _{i=1}^{r} ( \log ({\kappa_i}) - {\kappa_i} +1 )$. Note that $\log ({\kappa} ) - {\kappa} +1$ is strictly decreasing for all ${\kappa} \ge 1$. As can be seen, $g({\bm{V}^{}}^{\prime})$ is minimized for ${\kappa_i} = \max \{ 1,\lambda _i^{\prime} \}$ for $i \in [r]$, and $\lambda _1^{\prime} \ge \cdots \ge \lambda _r^{\prime}$ are the top $r$ eigenvalues of ${{\bm {R}}^{}}^{\prime}$. 

The best option for ${{\bm{\zeta}}_i}$ is given by ${{\bm{\zeta}}_i} = 0$ when ${\kappa_i}=1$. When ${\kappa_i} > 1$, ${{\bm{\zeta}}_i}$ is an eigenvector of ${{\bm {R}}^{}}^{\prime}$ with eigenvalue $\lambda _i^{\prime}$ and we have that ${{\bm{\zeta}}_i}^T {{\bm{\zeta}}_i} = \max \{ 1, \lambda _i^{\prime} \} -1$. Finally, we note that $\min _{{\bm {\bm{P}^{}} }\succeq \epsilon \bm {I}, {\bm{V}^{}}} h({ \bm {\bm{P}^{}} }, {\bm{V}^{}}) = \min _{{\bm {\bm{P}^{}} }\succeq \epsilon \bm {I}}\left\{ \min _{{\bm{V}^{}}}  h({\bm {\bm{P}^{}} }, {\bm{V}^{}})\right\}$. In fact, the method of minimizing the objective function with respect to $\bm{V}$ with $\bm{P}$ held fixed, is based on the classical work of the maximum likelihood factor analysis [30]. More specifically, since $\partial h(\bm {V}, \bm {P} )/{\partial \bm {P}} = \mathrm {diag}(\bm\Sigma ^{-1}(\bm\Sigma-\bm R)\bm\Sigma^{-1})$, the expression for the $i$th entry of $\partial h(\bm {V}, \bm {P} )/{\partial \bm {P}}$ is given by ${(\bm \Sigma _{(i,i)} - \bm R _{(i,i)})}/{p_i^2}$. We consider two cases, depending upon whether an optimal solution ${\hat{p }}_{i}$ satisfies: ${\hat{p }}_{i} > \epsilon$ or ${\hat{p }}_{i} = \epsilon$. If ${\hat{p }}_{i} > \epsilon$, then ${\partial h( \bm {V}, \bm P)}/{\partial {p _i}} = 0$, hence $\bm \Sigma _{(i,i)} = \bm R _{(i,i)}$ implies that $\bm R _{(i,i)} \ge {\hat{p }}_{i} > \epsilon$. Otherwise, if ${\hat{p }}_{i} = \epsilon$, then ${\partial h( \bm {V}, \bm P)}/{\partial {p _i}} \le 0$, this leads to $\bm R _{(i,i)} \ge {\hat{p }}_{i}=\epsilon$. Let $\bm{{\hat{{\bm{P}^{}} }}}$ be a solution of problem (P2), Appendix A demonstrates that any solution of problem (P2) is bounded. We get formulation problem (P2) by substituting the value of $\bm{V}$ that minimizes the inner minimization problem above into the objective function $h(\bm{{\bm{P}^{}} },{\bm{V}^{}})$, which proves Proposition 1. $\hfill\blacksquare$ 

However, problem (P2) is still non-convex due to the non-convex objective function, and thus still difficult to solve. Then, to this end, by introducing a new variable $\bm{\Gamma}\triangleq \bm{P}^{-1}= {\rm {diag}}( \gamma_1, \ldots ,\gamma_{\phi} )$, the following Proposition 2 shows that the objective function in problem (P2) can be expressed as a difference of two simple convex functions. 

\emph{Proposition 2: Define $\bm{\gamma} \triangleq ( \gamma_1, \ldots ,\gamma_{\phi} )$ as the new optimization variable, and $\bar \lambda ^{\prime}_{i}$, $i \in [r]$ are the top $r$ eigenvalues of $\bar {{\bm {R}}^{}}^{\prime} \triangleq \bm{\Gamma}^{\frac{1}{2}} {{\bm {R}}^{}} \bm{\Gamma}^{\frac{1}{2}}$, problem (P2) can be reformulated as the following optimization problem}
\begin{align}  
{(P3):}~~~\hbox {min}~~~&\displaystyle f(\bm{\gamma})\triangleq f_1(\bm{\gamma})-f_2(\bm{\gamma})\nonumber\\&=\sum _{i=1}^{\phi} ( -\log \gamma_i + {\varsigma _{i}}\gamma_i ) + e_r(\bar\lambda _i^{\prime}),\nonumber \\ \rm {s. t.}~~~&\bm {0} \prec \bm{\Gamma}={\rm {diag}}(\gamma_1, \ldots , \gamma_{\phi}) \preceq \frac{1}{\epsilon } \bm {I} \nonumber. 
\end{align}
where $\varsigma _i =\mathrm {max} \left\{ 0,1-\tfrac{1}{\lambda _i^{\prime}}\right\}  \text {if } 1 \le i \le r$, otherwise $\varsigma _i =0$. $\varsigma _i$ is widely used to computer the subgradients of the spectral functions [37] and particularly presented in the following Section IV-A. So far, problem (P3) is an instance of the well-known framework used for optimization of difference of convex problems [34] and can be effectively solved the convex concave procedure [35].

\emph{Proof:} Please refer to Appendix B. 
$\hfill\blacksquare$ 

Now, we can adopt a sequential linearization approach in which we linearize the function $f_{2}(\bm{\gamma})$ at each iteration, keeping $f_{1}(\bm{\gamma})$ unchanged, and solve the convex problem that results, also known as optimization of difference of convex problems [34] or convex concave procedure [35]. At last, based on the above analysis, we can start to introduce the complete detection framework of our JSTS-based method in the following Section. 

\section{Complete Detection Framework of JSTS-based method}

In this section, we present the complete detection framework of our JSTS-based method, including \emph{Stage 1: JSTS-based feature extraction} and \emph{Stage 2: sequential change frame detection}. Then, we also give the corresponding computation complexity analysis. We emphasize that we use the joint space-time sparsity to implement more robust UAJ detection and adopt a real-time sequential manner without assuming any attacker’s information. 

\subsection{JSTS-based Feature Extraction}

We use $\bm{\gamma}^{(m)}$ to denote the value of $\bm{\gamma}$ at the $m$th iteration, and linearize $f_{2}(\bm{\gamma})$ at $\bm{\gamma}^{(m)}$ with $f_{1}(\bm{\gamma})$ unchanged to obtain a convex approximation of $f(\bm{\gamma})$, denoted by $f(\bm{\gamma}) \approx f_1(\bm{\gamma})- ( f_2 ( \bm{\gamma}^{(m)}) + \langle \bm{\nabla }_{m},\bm{\gamma}-\bm{\gamma}^{(m)} \rangle ) \triangleq F(\bm{\gamma}; \bm{\gamma}^{(m)})$, where $\bm{\nabla }_{m}$ is a subgradient of $f_2 (\bm{\gamma} )$ at $\bm{\gamma}^{(m)}$ and $\left\langle \cdot , \cdot \right\rangle$ is the usual trace inner product. Next, we compute the subgradient of $f_2 (\bm{\gamma} )$ to ensure a satisfying approximation. Note that the gradient and subgradient of $f_1 (\bm{\gamma} )$ are the same since $f_1 (\bm{\gamma} )$ is differentiable. The main difficulty lies in the fact that $f_2 (\bm{\gamma} )$ is not differentiable, we first establish that ${\widetilde{H}}_{r}({\bm{o}_{r}}) \triangleq - H_{r}({\bm{o}_{r}})=-\sum \limits _{i=1}^{\phi}w_{i} g({{o}_{r,i}})=-\sum \limits _{i=1}^{\phi}w_{i} ( \log ( \max \{1,{{o}_{r,i}} \}) - \max \{1,{{o}_{r,i}}\} +1 )$ and then according to the differentiability of spectral functions as [37], the subgradient of ${\widetilde{H}}_{r}({\bm{o}_{r}})$ can be derived as $\partial {\widetilde{H}}_{r}({\bm{o}_{r}}) = -\sum _{i=1}^{\phi} {{w}}_{i} \nabla g({{o}_{r,i}})$, where $\nabla g({{o}_{r,i}}) $ is the gradient of $g({{o}_{r,i}})$ and $w_{i}$ is a minimizer of $g({{o}_{r,i}})$. Note that the function ${o}_{r,i} \mapsto \log ( \mathrm {max}\{1,{o}_{r,i} \}) -\mathrm {max} \{1,{o}_{r,i}\} +1$ is decreasing on ${o}_{r,i} \ge 0$. Hence the sum $\sum _{i=1}^{\phi}w_{i} ( \log ( \mathrm {max}\{1,{o}_{r,i} \}) -\mathrm {max} \{1,{o}_{r,i}\} +1 )$ will be minimized for a choice: $w_{i} = 1$ whenever ${o}_{r,i}$ is one of the top $r$ elements among ${o}_{r,1}, \ldots , {o}_{r,\phi}$; and $w_{i}=0$ for all other choices of $i \in [\phi]$. Let $\bm{ \Gamma } ^{\frac{1}{2}} {\bm {R}} \bm{\Gamma } ^{\frac{1}{2}} = \bm { U}_A\mathrm {diag}(\lambda _{1}^{\prime},...,\lambda _{p}^{\prime})\bm {U }_A^{{T} } $ be the eigen decomposition of ${{\bm{\Gamma }}} ^{\frac{1}{2}} {\bm {R}} \bm{\Gamma } ^{\frac{1}{2}}$, according to the properties of subgradients of spectral functions [37], $\partial f_{2}(\bm{\gamma })$ can be written as $\partial f_2( \bm{\gamma }) = \mathrm {diag}( \bm{\Gamma } ^{-\frac{1}{2}} \bm {U}_A\bm {D}_A\bm {U}_A ^{{T} } \bm{\Gamma }^{\frac{1}{2}} {\bm {R}})$, where $\bm {D}_A = \mathrm {diag}( \varsigma _1,...,\varsigma_{\phi})$ with $\varsigma _i =\mathrm {max} \left\{ 0,1-\tfrac{1}{\lambda _i^{\prime}}\right\}  \text {if } 1 \le i \le r$, otherwise $\varsigma _i =0$. By using the above subgradient $\nabla g({{o}_{r,i}}) $, $\bm{\gamma}^{(m+1)}$ can be calculated as 
\begin{align}
\bm{\gamma}^{(m+1)} = \mathop {\hbox {arg min}}\limits _{\frac{1}{\epsilon } \ge \gamma_i > 0, i \in [\phi]}\sum _{i=1}^{\phi} ( - \log \gamma_i + {\varsigma _{i}}\gamma_i - \nabla _{m,i}\gamma_i) , 
\end{align}
where $\nabla _{m,i}$ is the $i$th coordinate of $\nabla _{m} g({{o}_{r,i}})$, and given by $\nabla _{m} g({{o}_{r,i}}) = \min \{ 0, \frac{1}{{{o}_{r,i}}} - 1 \}$, and $\nabla _{m} g({{o}_{r,i}})=0$ for all $m \ne i$. Then, the $i$th entry of $\bm{\gamma}^{(m+1)}$ can be derived as 
\begin{align} 
{\gamma}^{(m+1)}_{i} = \min \left\{ \frac{1}{{\varsigma _{i}} - \nabla _{m,i}} ,\frac{1}{\epsilon } \right\} ~~~ \text {for}~~~ i \in [\phi]. 
\end{align}

The updates will keep going till the stopping criterion $\Vert \bm{\gamma}^{(m+1)} - \bm{\gamma}^{(m)}\Vert _{2} < \varepsilon \Vert \bm{\gamma}^{(m)}\Vert _{2}$ is met, where $\varepsilon >0$ is small positive number determining the accuracy of the JSTS-based feature extraction.

\subsection{Sequential Change Frame Detection}

After extraction of the JSTS-based feature, we can perform sequential change frame detection in a real-time manner to quickly detect UAJ. 

1) Sequential process: At each time when a new sample observation is obtained, the detector makes a decision based on all the sample observations collected at hand. Combined with UAJ detection in this work, if it indicates that no change in the joint space-time sparsity has occurred, then the detector moves to the next frame instant, collecting new samples and making a new decision. We can see that with sequential process, our detection method works in a real-time manner and does not rely on the prior knowledge of the attackers. 

2) Change frame detection: We use $\left\{ {{\bm{o}^{( {l-1,t} )}}} \right\}_{t = 1}^{{T_s}}$ and $\left\{ {{\bm{o}^{( {l,t} )}}} \right\}_{t = 1}^{{T_s}}$ to denote the history monitoring samples in the $(l-1)$th frame and the newly arriving samples in the $l$th frame, respectively. The support size of ${\tau ^{(l)}}$, i.e., the JSTS-based feature, can be calculated by ${\tau ^{( l )}} = \sum\limits_{\zeta = 1}^{\phi} \delta_D ({[\bm{\gamma}_{[l]}]_\zeta \ne 0})$, and ${\tau ^{(l-1)}}$ of the history monitoring samples in the $(l-1)$th frame can be obtained in the same way. In this paper, in order to detect if the newly arriving samples $\left\{ {{\bm{o}^{( {l,t} )}}} \right\}_{t = 1}^{{T_s}}$ in $l$th frame is anomalous due to the access jamming, we proposed to check if the joint space-time sparsity has experienced a big change. To capture the change of the joint space-time sparsity based feature between the monitoring samples and the newly arriving samples, we define a metric $c\triangleq 1 - \left| {\frac{{{\tau ^{( l )}} - {\tau ^{( {l - 1} )}}}}{{{\tau ^{( {l - 1} )}}}}} \right|$. A smaller $c$ corresponds to larger changes of the joint space-time sparsity. Accordingly, if the newly arriving samples $\left\{ {{\bm{o}^{( {l,t} )}}} \right\}_{t = 1}^{{T_s}}$ result in a large change of the joint space-time sparsity based feature and thus $c$ is lower than the pre-defined detection threshold $\delta$, we will claim that the access jamming exists. Based on the probability theory and statistics, we use the cumulative distribution function to help to set up $\delta$.

\begin{algorithm}[t]
\caption{JSTS-based method for UAJ detection}
\label{JSTS}
\begin{algorithmic}[1]
\STATE \textbf{Input:} Newly arriving signals in the $l$th frame $\left\{ {{\bm{o}^{( {l,t} )}}} \right\}_{t = 1}^{{T_s}}$; History signals in $(l-1)$th frame $\left\{ {{\bm{o}^{( {l-1,t} )}}} \right\}_{t = 1}^{{T_s}}$; Stopping criterion $\varepsilon$; Detection threshold $\delta$. 

\STATE $~~~$\textbf{Stage 1:} JSTS-based feature extraction.

\STATE $~~~~$Compute the JSTS-based feature on newly arriving signals in the $l$th frame to get ${\tau ^{(l)}}$.

\STATE $~~~~~~~~$\textbf{for} $t=1$ \textbf{to} $T_s$ \textbf{do}

\STATE $~~~~~~~~$${\bm{o}^{( {l,t} )}}$ is expanded as a real one $\bm{o}_r^{( {l,t} )}$.

\STATE $~~~~~~~~$\textbf{end for}

\STATE $~~~~~~~~$Build the equivalent measure vector $\bm{o}_r^{( l )}$;

\STATE $~~~~~~~~$\textbf{Initialize} $m=1$

\STATE $~~~~~~~~~~$\textbf{Repeat:}

\STATE $~~~~~~~~~~~~$Compute ${\nabla _m}$ according to $\nabla g({{o}_{r,i}}) $;

\STATE $~~~~~~~~~~~~$Update $\bm {\gamma } _{[l]}^{( m+1 )}$ according to (6) and (7);  

\STATE $~~~~~~~~~~$\textbf{Until:} ${\left\| \bm {\gamma } _{[l]}^{( m+1 )} - \bm {\gamma } _{[l]}^{( m )}\right\|_2} < \varepsilon {\left\| \bm {\gamma } _{[l]}^{( m )} \right\|_2}$.

\STATE $~~~~~~~~$Compute the support size of ${\tau ^{(l)}}$ according to ${\tau ^{( l )}} = \sum\limits_{\zeta = 1}^{\phi} \delta_D ({[\bm{\gamma}_{[l]}]_\zeta \ne 0})$;

\STATE $~~~~$Compute the JSTS-based feature on history signals in $(l-1)$th frame to get ${\tau ^{(l-1)}}$. 

\STATE $~~~~~~~~$Compute the support size of ${\tau ^{(l-1)}}$;

\STATE $~~~$\textbf{Return:} The list of the JSTS-based feature $[{\tau ^{(l-1)}},{\tau ^{(l)}}]$.

\STATE $~~~$\textbf{Stage 2:} Sequential change frame detection.

\STATE $~~~~$Compute $c \triangleq 1 - \left| {\frac{{{\tau ^{( l )}} - {\tau ^{( {l - 1} )}}}}{{{\tau ^{( {l - 1} )}}}}} \right|$. 

\STATE $~~~~~~~~$\textbf{if} $c > \delta $ \textbf{then}

\STATE $~~~~~~~~~~$${\bm{o}_r^{( {l} )}}$ is normal;

\STATE $~~~~~~~~~~$Save ${\tau ^{(l)}}$ for the next detection;

\STATE $~~~~~~~~$\textbf{else} 

\STATE $~~~~~~~~~~$${\bm{o}_r^{( {l} )}}$ is under UAJ;

\STATE $~~~~~~~~~~$Update ${\tau ^{(l)}}={\tau ^{(l-1)}}$ for the next detection;

\STATE $~~~~~~~~$\textbf{end if} 

\STATE $~~~$\textbf{Return:} Detection result.

\end{algorithmic}
\end{algorithm}

To sum up, combining Stage 1 (the JSTS-based feature extraction) and Stage 2 (the sequential change frame detection), UAJ detection can be efficiently carried out as outlined in Algorithm 1. Algorithm 1 shows the complete framework of our JSTS-based method. The JSTS-based feature extraction is shown in Stage 1. The calculation of subgradients ${\nabla _m}$ and convex approximation of $f( { \bm {\gamma }_{[l]}} )$ are the core of this stage. Then, if the newly arriving samples result in a large change of the JSTS-based feature and thus $c$ is lower than the pre-defined detection threshold $\delta$, we will claim that UAJ exists, and only update the support size of sparsity utilized for the next detection, as shown in Stage 2.

\subsection{Computational Complexity Analysis}

In this part, we will present the complexity analysis of the proposed JSTS-based detection method and discuss the computational scalability to large problems. Depending on the relative sizes of the number of time slots $T_s$ and the data dimension $N$, we present the complexity analysis for the following three cases: 

\emph{Case I ($T_s>N$):} The main computational cost of the JSTS-based detection method is the computation of a subgradient of $f_2 (\bm{\gamma} )$ which needs a low-rank eigen decomposition of ${{\bm {R}}^{}}^{\prime}$. When $N$ is small relative to $T_s$, it is simple to form and work with matrix ${{\bm {R}}^{}}^{\prime}$. We can creat ${{\bm {R}}^{}}$ from $\bm{o}_{r}$ offline and compute ${{\bm {R}}^{}}^{\prime}$ from ${{\bm {R}}^{}}$, which respectively cost $O({T_s^2}{N})$ and $O(N^2)$. As for the direct low-rank eigen-decomposition of ${{\bm {R}}^{}}^{\prime}$ costs $O(N^3)$ and can not apply to large-scale problems. 

\emph{Case II ($T_s \ll N$):} In several delay-sensitive applications, $T_s$ is usually less than $N$ in order to achieve a timely detection. In such circumstances, $\bm{\Gamma}^{\frac{1}{2}} {{\bm {R}}^{}} \bm{\Gamma}^{\frac{1}{2}} = ( \tfrac{1}{\sqrt{T_s}}\bm {o}_r\bm{\Gamma}^{\frac{1}{2}})^T (\tfrac{1}{\sqrt{T_s}}\bm {o}_r\bm{\Gamma}^{\frac{1}{2}})$ can be gained by a  singular value decomposition (SVD) of $\tfrac{1}{\sqrt{T_s}}\bm {o}_r\bm{\Gamma}^{\frac{1}{2}}$, which costs $O(T_s^2N)$. Thus, it will cost $O(T_s^2N)$ to compute a rank $r$ eigen decomposition of ${{\bm {R}}^{}}^{\prime}$, which is linear in $N$ if $T_s \ll N$. 

\emph{Case III (both $T_s$ and $N$ are large):} When both $T_s$ and $N$ are large, the above direct SVD method will become computationally expensive. For large-scale low-rank SVD decompositions, we need to rely on some approximate techniques. By using the approximate rank $r$ eigen decompositions in [38], the computational complexity can be quickly reduced to $O(N^2 r)$ for $r \ll N \approx T_s$.

As for the conventional batch detection, the energy characteristic (EC) based method [15] needs to calculate the norm of the received signal, the computing complexity of which is $O({T_s^2}{N^2})$. In addition, the second-order statistics (SOS) based method [19] needs to calculate the inverse of a $2T_sN$-dimensional matrix, the computing complexity of which is $O({T_s^3}{N^3})$. By contrast, our method is more suitable for UAJ detection in practice.

\section{Numerical Results}

In this section, some numerical examples are performed to investigate the performance of the proposed JSTS-based method for UAJ detection, and provided to verify our theoretical results. We consider $K = 2000$ potential devices that attempt to access a mission-critical mMTC network. The MTDs and UAJ attackers are randomly distributed in a circular region with the BS located at the center. The distances from the MTDs and UAJ attackers to the BS, i.e., $D_k^{},k = 1, \cdots ,K$ and $D_{A,j}^{},j = 1, \cdots ,J$, are randomly distributed in the region $[40\rm m,800 \rm m]$ and $[60 \rm m, (D_{max}) \rm m]$, respectively. The large-scale fading from the MTDs and UAJ attackers to the BS are set to be ${\beta _k} = {L_o}D_k^{ - \alpha }$ and ${\beta _{A,j}} = {L_o}D_{A,j}^{ - \alpha }$, where $\alpha  = 4$ is the path loss exponent and ${L_o} =  - 45$dB denotes the shadowing effect. We assume that all MTDs have the same transmit power, i.e., $P_k^{} = P = 20$dBm, and the jamming power is also set to be the same for all UAJ attackers, i.e., $P_{A,j}^{} = P_{UAJ}$. We consider a 20 MHz channel with noise floor of  $-101$ dBm. The length of spreading sequence is $N  = 50$. In each frame, the ensemble of sparse signals $d_k^{( {l,t} )}$ are randomly generated from the QPSK symbol set for the consecutive $T_s=7$ time slots in each frame. We set the common part in any two adjacent time slots for each frame satisfies ${\eta ^{( l )}} = 0.3,\forall l$. Unless specified, we set $\mu=0.05$, $\rho=0.4$, $\varepsilon=10^{-3}$, $J=8$, $D_{\rm {max}}=800$ and $P_{UAJ}=20$dBm. As a note, to compare the detection accuracy of different methods, with other parameters fixed, we vary the access probability $\rho$, the number of UAJ attackers $J$, the maximum distance of UAJ attackers $D_{\rm {max}} $, and the transmit power of UAJ attackers $P_{UAJ}$. In addition, the EC-based method [15], the SOS-based method [19], the ML-based factor analysis [32] and the EM-based factor analysis [33] are considered for comparison. 

\begin{figure}[t]
  \centering
  \includegraphics[width=2.8 in]{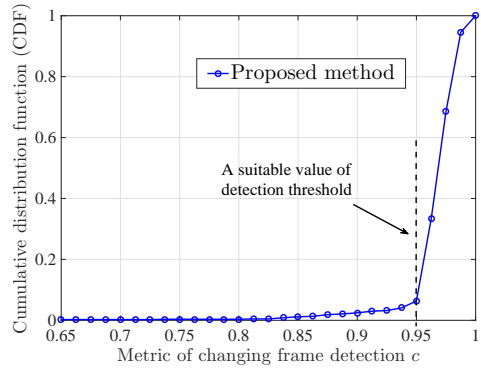}\\
  \caption{CDF result of detection threshold $\delta$.}
\end{figure}

\begin{figure}[t]
  \centering
  \includegraphics[width=2.8 in]{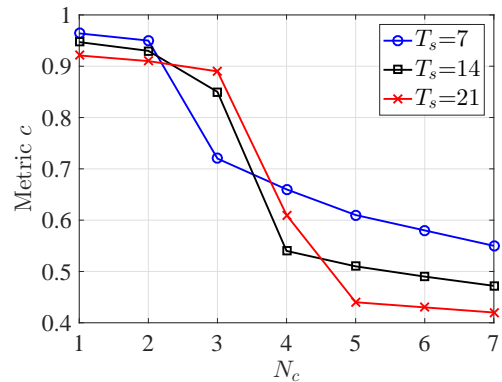}\\
  \caption{Metric $c$ versus $N_c$.}
\end{figure}

According to the proposed JSTS-based method, if the joint space-time sparsity changed abruptly, i.e., the specific value $c \triangleq 1 - \left| {\frac{{{\tau ^{( l )}} - {\tau ^{( {l - 1} )}}}}{{{\tau ^{( {l - 1} )}}}}} \right|$, as defined in Section IV-B, is lower than the pre-defined detection threshold $\delta $, we will determine that UAJ occurs. We set the detection threshold $\delta$ based on the probability theory and statistics. Specifically, we use the cumulative distribution function (CDF) to help to set up $\delta$. Fig. 3 shows the CDF result of detection threshold $\delta$ by using the history signal in previous frame. In order to obtain the standard of change frame detection, we assume UAJ does not exist during the training phase of detection threshold acquisition. As nearly all the $\delta$ is larger than 0.95, thus we set $\delta=0.95$ under the above settings. In Fig. 4, we evaluate the impact on the JSTS-based feature by different number of consecutive time slots activated by the attackers. We use the metric $c$ to capture the change of the JSTS-based feature between the monitoring samples in the previous frame and the newly arriving samples in the current frame, and $N_c$ to denote the number of consecutive time slots activated by the attackers. As can be observed, for three cases with different slot configuration, a sudden change in the JSTS-based feature can be found, even if each attacker sends the spreading sequences in a small number of consecutive time slots which has relatively small proportion in the whole slot configuration. For example, there exists an obvious abrupt change in the JSTS-based feature when $N_c=3$ ($N_c=4$) for $T_s=7$ ($T_s=14$ or 21). Therefore, the proposed UAJ detection is sensitive and reliable, which has broad application prospect in practice.

\begin{figure}[t]
  \centering
  \includegraphics[width=2.8 in]{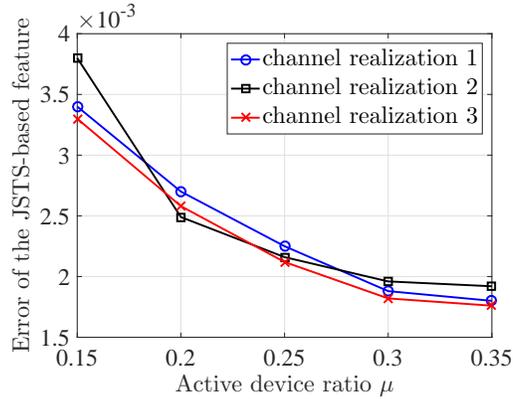}\\
  \caption{The quality of the solution of the proposed JSTS-based method}
\end{figure}

\begin{figure}[t]
  \centering
  \includegraphics[width=2.8 in]{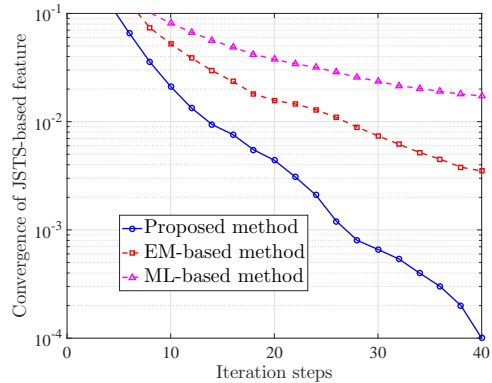}\\
  \caption{The comparison of convergence of different methods.}
\end{figure}

In Fig. 5, we investigate the quality of the solution of the JSTS-based feature extracted by the proposed method for 3 random channel realizations. Specifically, we plot the error of the JSTS-based feature between the estimation result from the proposed method with the actual value from the synthetic data. As can be observed, the proposed method achieves almost the same estimation result as the actual value for different active device ratio $\mu$. Fig. 6 illustrates the convergence of the value of detection feature versus iteration steps through the proposed JSTS-based method and the two factor analysis methods. We can observe that the JSTS-based method can guarantee convergence and requires significantly less iterations than the ML-based method and the EM-based method. Although the proposed method takes a certain number of iterations to reach convergence, the running speed of the whole process is fast due to low computational complexity.

\begin{figure}[t]
  \centering
  \includegraphics[width=2.8 in]{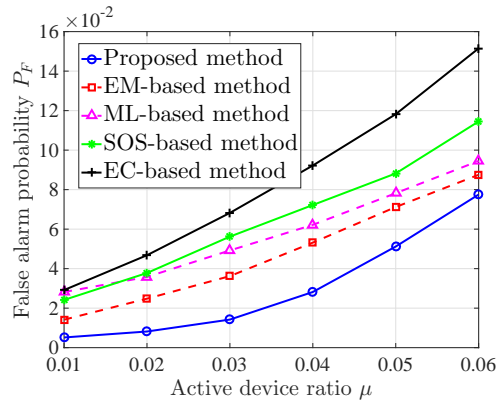}\\
  \caption{The comparison of false alarm probability of different methods.}
\end{figure}

\begin{figure}[t]
  \centering
  \includegraphics[width=2.8 in]{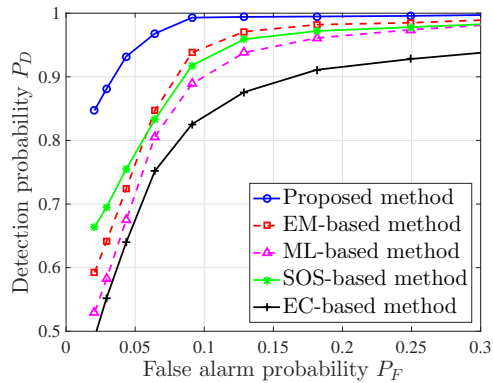}\\
  \caption{The comparison of ROC of different methods.}
\end{figure}

To demonstrate the performance of the proposed JSTS-based method, we make a comparison of different methods in terms of false alarm probability $P_F$ in Fig. 7 and the receiver operating characteristic (ROC) curves in Fig. 8. Firstly, Fig. 7 depicts the comparison result of $P_F$ of different methods for different active device ratio $\mu$. Because the active device ratio $\mu$ is closely related to the space sparsity in mMTC, we first make a comparison of different methods with varying active device ratio $\mu$. It can be seen from Fig. 7 that $P_F$ can be controlled well in the proposed method compared to other methods. In addition, $P_F$ increases as $\mu$ becomes larger for all methods. This is because higher $\mu$ makes more devices connected to the BS, which leads the detection features under the space sparsity constraints to cause the erroneous judgement easily. Secondly, for the sake of fairness, we make a performance comparison of different methods with a series of values at fixed $P_F$. To accomplish this, we evaluate the detection performance of different methods by the receiver operating characteristic (ROC) curves in Fig. 8. As we can see, the proposed JSTS-based method shows a better detection performance than other methods for a given false alarm probability $P_F$. For example, detection probability $P_D$ of our JSTS-based method reaches above 0.95 when $P_F=0.05$. In contrast, the detection accuracy of the competing methods are still far behind the proposed method, which indicates they are not suitable for UAJ detection.

\begin{figure}[t]
  \centering
  \includegraphics[width=2.8 in]{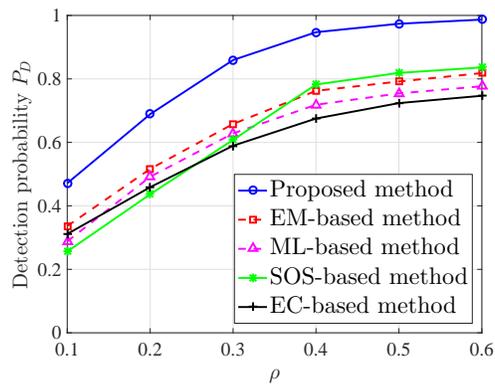}\\
  \caption{Detection accuracy comparisons with respect to the access probability $\rho$.}
\end{figure}

\begin{figure}[t]
  \centering
  \includegraphics[width=2.8 in]{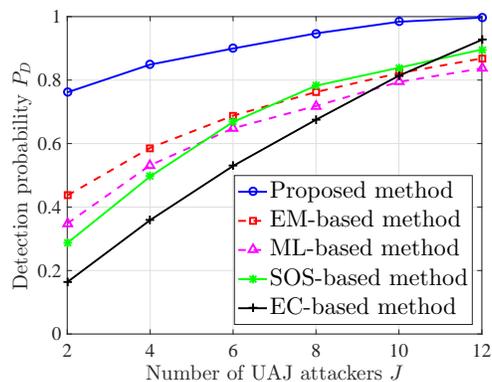}\\
  \caption{Detection accuracy comparisons with respect to the number of UAJ attackers.}
\end{figure}

In Fig. 9, we compare the detection performances of different methods under different access probability $\rho$. We set $J=8$, $D_{\rm {max}}=800$, and $P_{UAJ}=20$dBm. As can be observed, the detection accuracy increases with increasing $\rho$. This is due to the fact that the damage on the time sparsity caused by UAJ increased with increasing collision probability, which further leads to an obvious change of the JSTS-based feature. Moreover, there is always a significant performance difference between the proposed method and other methods. Although the same JSTS-based feature are used to detect UAJ in the ML-based method and the EM-based method, they are prone to get stuck in suboptimal local solution of the feature extraction and thus their detection performance are not very satisfactory. In Fig. 10, we illustrate the detection performances of different methods versus the number of UAJ attackers $J$, where we set $\rho=0.4$, $D_{\rm {max}}=800$, and $P_{UAJ}=20$dBm. Fig. 10 reveals that the detection accuracy increases with increasing $J$ for all the methods, and the proposed JSTS-based method always performs better in detection accuracy than other methods under the same number of UAJ attackers.

\begin{figure}[t]
  \centering
  \includegraphics[width=2.8 in]{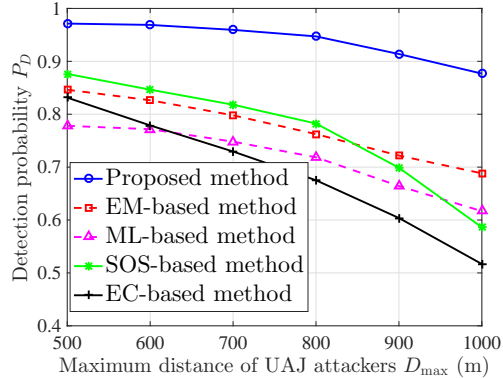}\\
  \caption{Detection accuracy comparisons with respect to the maximum distance of UAJ attackers.}
\end{figure}

\begin{figure}[t]
  \centering
  \includegraphics[width=2.8 in]{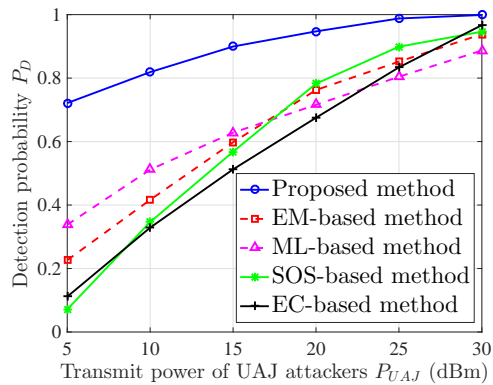}\\
  \caption{Detection accuracy comparisons with respect to the transmit power of UAJ attackers.}
\end{figure}

We observe the effect of the maximum distance between the BS and the attackers $D_{\rm {max}}$ in Fig. 11. We set $\rho=0.4$, $J=8$, and $P_{UAJ}=20$dBm. For all methods, we observe the detection probability decreases with the increasing of $D_{\rm {max}}$. Compared with these competing methods, even if the $D_{\rm {max}}$ becomes larger, the detection performance of our JSTS-based method has a small decline and appears to be robust enough to widely distributed attackers. In Fig. 12, we plot the detection probability versus the transmit power of UAJ attackers $P_{UAJ}$, where we set $\rho=0.4$, $J=8$, and $D_{\rm {max}}=800$. As we can see, with the increase of $P_{UAJ}$, UAJ attackers will be easily distinguished by the BS with the help of the JSTS-based method. This is because the JSTS-based method utilizes difference of convex optimization for obtaining feasible solutions to the problem of the time sparsity extraction, and a higher attacking power will produce a larger deviation the time sparsity estimation procedure. In addition, we note that within a vast range value of $P_{UAJ}$, the JSTS-based method always outperforms these competing methods.

\section{Conclusion}
\label{Sec:Conclusion}

In this paper, we have investigated the design of a detection method for the uplink access jamming (i.e., UAJ) in the mission-critical mMTC. In order to obtain a timely and reliable detection result, we have extracted a new feature based on the joint space-time sparsity, i.e., the JSTS-based feature, which can reflect the intrinsic properties of mMTC and is very sensitive to UAJ. On this basis, we have performed the JSTS-based detection in a sequential manner, the detector made a decision through the abrupt change in the JSTS-based feature without knowing the accurate prior information of the attackers. Numerical results have shown the excellent detection performance of our proposed method.

\appendix

\subsection{The proof of the bounds on an optimal solution of (P2).}
{

First of all, by setting $\partial {h(\bm{P },\bm {V})}/{\partial \bm {V}}$ to zero, and applying the S. W. formula on $ (\bm{P }+ \bm {V}\bm {V}^T )^{-1}$ [36], we have that ${\bm {V}} = {\bm {R}} \bm{P }^{-1}{\bm {V}}( \bm {I} + {\bm {V}}^{T} \bm{P }^{-1}{\bm {V}} ) ^{-1}$. In addition, substituting (5) into ${\bm {R}}{\bm{\Sigma }}^{-1}$, we have 
\begin{align} 
{\bm {R}}{\bm{\Sigma }}^{-1}= & {} {\bm {R}}\bm{P }^{-1} - ( {\bm{\Sigma }} - \bm{P }) \bm{P }^{-1} ~~~~~~~~~\nonumber \\
= & {} {\bm {R}}\bm{P }^{-1} - {\bm{\Sigma }}\bm{P }^{-1} + \bm {I}.
\end{align}

Moreover, $\bm{\Sigma }^{-1} ( \bm{\Sigma }- {\bm {R}}) \bm{\Sigma }^{-1}$ can be calculated by
\begin{align} 
\bm{\Sigma }^{-1} ( \bm{\Sigma }- {\bm {R}}) \bm{\Sigma }^{-1}= & {} \bm{\Sigma }^{-1} - \bm{\Sigma }^{-1} ({\bm {R}} \bm{\Sigma }^{-1})\nonumber \\= & {} - ( \bm{P }^{-1}{\bm {R}} -\bm{P }^{-1}\bm{\Sigma }+\bm {I} ) \bm{P }^{-1} + \bm{P }^{-1} \nonumber \\= & {} \bm{P }^{-1} ( \bm{\Sigma }- {\bm {R}}) \bm{P }^{-1}. 
\end{align}

Because $\partial h({\bm {V}}, \bm{P })/{\partial \bm{p}} = \mathrm {diag}(\bm{\Sigma }^{-1} ( \bm{\Sigma }- {\bm {R}}) \bm{\Sigma }^{-1})$, combing (9), the $i$th entry of $\partial h({\bm {V}}, \bm{P })/{\partial \bm{p }}$ can be written as ${(\bm \Sigma _{(i,i)} - \bm R _{(i,i)})}/{p_i^2}$. We consider two cases, depending upon whether an optimal solution ${\hat{p }}_{i}$ satisfies: ${\hat{p }}_{i} > \epsilon$ or ${\hat{p }}_{i} = \epsilon$. If ${\hat{p }}_{i} > \epsilon$, then ${\partial h( \bm {V}, \bm P)}/{\partial {p _i}} = 0$, hence $\bm \Sigma _{(i,i)} = \bm R _{(i,i)}$ implies that $\bm R _{(i,i)} \ge {\hat{p }}_{i} > \epsilon$. Otherwise, if ${\hat{p }}_{i} = \epsilon$, then ${\partial h( \bm {V}, \bm P)}/{\partial {p _i}} \le 0$, this leads to $\bm R _{(i,i)} \ge {\hat{p }}_{i}=\epsilon$. This completes the proof.
 
}

\subsection{The proof of Proposition 2}
{

To prove ${\widetilde{H}}_{r}({\bm{o}_{r}})\triangleq - H_{r}({\bm{o}_{r}})=-\sum \limits _{i=1}^{\phi}w_{i} g({{o}_{r,i}})$ is concave on ${{o}_{r}}$, where $H_{r}({\bm{o}_{r}})$ is a linear functional, $H_{r}({\bm{o}_{r}})$ can first be written as 
\begin{align}
\sum \limits _{i=1}^{\phi} w_{i}( \log ( \max \{1,{{o}_{r,i}} \}) - \max \{1,{{o}_{r,i}}\} +1 ). 
\end{align}

Because the scalar function $( \log ( \max \{1,{{o}_{r,i}} \}) - \max \{1,{{o}_{r,i}}\} +1 )$ is decreasing on ${o}_{r,i}$, thus the sum $\sum \limits _{i=1}^{\phi} w_{i}( \log ( \max \{1,{{o}_{r,i}} \}) - \max \{1,{{o}_{r,i}}\} +1 )$ will be minimized for a choice: $w_{i} = 1$ whenever ${o}_{r,i}$ is one of the top $r$ elements among ${o}_{r,1}\cdots {o}_{r,\phi}$; $w_{i} = 0$  for all other choices of ${o}_{r,i}$. As a result, representation is justified. $( \log ( \max \{1,{{o}_{r,i}} \}) - \max \{1,{{o}_{r,i}}\} +1 )$  is concave for any ${o}_{r,i}$. So, for every fixed $w_{i} \ge 0$, the function $\sum \limits _{i=1}^{\phi} w_{i}( \log ( \max \{1,{{o}_{r,i}} \}) - \max \{1,{{o}_{r,i}}\} +1 )$ is concave on ${o}_{r,i}$. The linearity of the map $\bm{\gamma}$ to $\bm{\Gamma}^{\frac{1}{2}} {{\bm {R}}^{}} \bm{\Gamma}^{\frac{1}{2}}$ indicates that $f(\bm{\gamma})$ is concave on $\bm{\gamma}$. Let $\bm{ \Gamma } ^{\frac{1}{2}} {\bm {R}} \bm{\Gamma } ^{\frac{1}{2}} = \bm { U}_A\mathrm {diag}(\lambda _{1}^{\prime},...,\lambda _{p}^{\prime})\bm {U }_A^{{T} } $ be the eigen decomposition of ${{\bm{\Gamma }}} ^{\frac{1}{2}} {\bm {R}} \bm{\Gamma } ^{\frac{1}{2}}$, according to the properties of subgradients of spectral functions [37], $\partial f_{2}(\bm{\gamma })$ can be written as $\partial f_2( \bm{\gamma }) = \mathrm {diag}( \bm{\Gamma } ^{-\frac{1}{2}} \bm {U}_A\bm {D}_A\bm {U}_A ^{{T} } \bm{\Gamma }^{\frac{1}{2}} {\bm {R}})$, where, $\bm {D}_A = \mathrm {diag}( \varsigma _1,...,\varsigma_{\phi})$ with $\varsigma _i =\mathrm {max} \left\{ 0,1-\tfrac{1}{\lambda _i^{\prime}}\right\}  \text {if } 1 \le i \le r$, otherwise $\varsigma _i =0$. By using the above subgradient $\nabla g({{o}_{r,i}}) $, the value of $\bm{\gamma}$ at the $m$th iteration, denoted by $\bm{\gamma}^{(m)}$, can be obtianed. 

From convexity of $f_2( \bm{\gamma})$, we have that
\begin{align} 
f_{2}( \bm{\gamma }^{(m+1)}) \ge &f_{2}( \bm{\gamma }^{(m)})\nonumber \\&+\left\langle \partial f_{2}( \bm{\gamma }^{(m)}) , \bm{\gamma }^{(m+1)}-\bm{\gamma }^{(m)}\right\rangle ,
\end{align}
where $\left\langle \cdot , \cdot \right\rangle$ is the usual trace inner product. $\bm{\gamma}^{(m)}$ is the value of $\bm{\gamma}$ at the $m$th iteration and $\partial f_{2}( \bm{\gamma }^{(m)})$ denotes the subgradient of $\bm{\gamma}^{(m)}$. Since $f_1 (\bm{\gamma} )$ is differentiable, we have that
\begin{align} 
f_{1}( \bm{\gamma }^{(m)}) \ge &f_{1}( \bm{\gamma }^{(m+1)}) + \left\langle \bm{\gamma }^{(m)}-\bm{\gamma }^{(m+1)},\nabla f_{1}( \bm{\gamma }^{(m+1)}) \right\rangle \nonumber \\&+ \frac{\psi}{2}\Vert \bm{\gamma }^{(m+1)}-\bm{\gamma }^{(m)}\Vert ^{2}, 
\end{align}
where $\nabla f_{1}( \bm{\gamma }^{(m+1)})$ is the derivative of $f_{1}( \bm{\gamma }^{(m)})$ and $\psi$ denotes a coefficient of strong convexity for $f_{1}( \bm{\gamma }^{(m)})$. Combining (11) and (12), we further have that 
\begin{align} 
f_1( \bm{\gamma }^{(m+1)} ) -f_2 ( \bm{\gamma }^{(m+1)})\le &f_1( \bm{\gamma }^{(m)} ) -f_2 ( \bm{\gamma }^{(m)}) \nonumber \\&- \frac{\psi}{2}\Vert \bm{\gamma }^{(m+1)}-\bm{\gamma }^{(m)}\Vert ^{2}. 
\end{align}

Combining with the above characteristics, problem (P3) converges to a stationary point of the original non-convex problem, which proves Proposition 2.

}

\end{document}